\documentclass[11pt,english]{article}
\usepackage{mathptmx}
\usepackage[T1]{fontenc}
\usepackage[latin1]{inputenc}
\usepackage[letterpaper]{geometry}
\usepackage{array}
\usepackage{float}
\usepackage{amstext}
\usepackage{graphicx}
\usepackage{amssymb}

\makeatletter

\providecommand{\tabularnewline}{\\}

\newcommand{\lyxaddress}[1]{
\par {\raggedright #1
\vspace{1.4em}
\noindent\par}
}

\author{}
\date{}
\usepackage{cite}

\@ifundefined{showcaptionsetup}{}{%
 \PassOptionsToPackage{caption=false}{subfig}}
\usepackage{subfig}
\makeatother

\usepackage{babel}

\begin{document}

\title{Thickness-dependent spontaneous dewetting morphology of ultrathin
Ag films}

\author{$^{1}$H. Krishna, $^{2}$R. Sachan, $^{2}$J. Strader, $^{1}$C.
Favazza, $^{3}$M. Khenner and $^{2,4,5}$Ramki Kalyanaraman%
\thanks{Corresponding author, ramki@utk.edu%
}}

\maketitle

\lyxaddress{\begin{center}
$^{1}$Department of Physics, Washington University in St. Louis,
MO 63130\\
$^{2}$Department of Material Science and Engineering, University
of Tennessee, Knoxville, TN 37996\\
$^{3}$Department of Mathematics, Western Kentucky University,
Bowling Green, KY 42101\\
$^{4}$Department of Chemical and Biomolecular Engineering, University
of Tennessee, Knoxville, TN 37996\\
$^{5}$Sustainable Energy Education and Research Center, University
of Tennessee, Knoxville, TN 37996
\par\end{center}}
\begin{abstract}
We show here that the morphological pathway of spontaneous dewetting
of ultrathin Ag films on SiO$_{2}$ under nanosecond laser melting
is found to be film thickness dependent. For films with thickness
\emph{h} between $2\leq h\leq9.5$~nm, the morphology during the
intermediate stages of dewetting consisted of bicontinuous structures.
For films $11.5\leq h\leq20$~nm, the intermediate stages consisted
of regularly-sized holes. Measurement of the characteristic length
scales for different stages of dewetting as a function of film thickness
showed a systematic increase, which is consistent with the spinodal
dewetting instability over the entire thickness range investigated.
This change in morphology with thickness is consistent with observations
made previously for polymer films {[}\emph{A. Sharma et al, Phys.
Rev. Lett., v81, pp3463 (1998); R. Seemann et al, J. Phys. Cond. Matt.,
v13, pp4925, (2001)}{]}. Based on the behavior of free energy curvature
that incorporates intermolecular forces, we have estimated the morphological
transition thickness for the intermolecular forces for Ag on SiO$_{\text{2}}$.
The theory predictions agree well with observations for Ag. These
results show that it is possible to form a variety of complex Ag nanomorphologies
in a consistent manner, which could be useful in optical applications
of Ag surfaces, such as in surface enhanced Raman sensing.
\end{abstract}

\section{Introduction}

Silver (Ag) films and nanostructures have strong plasmonic activity
and consequently, are very useful in the chemical detection of species
via surface enhanced Raman scattering (SERS) \cite{quinten98,vanduyne07}.
It is known that the magnitude of the localized field enhancement
leading to increase in Raman scattering is very sensitive to the roughness
or asymmetry (aspect ratio) of the nanostructures \cite{fleischmann74}.
Therefore, controlling the morphology and nanostructure characteristics,
of metals like Ag, in a reliable and cost-effective manner is very
important towards further improving the sensitivity and selectivity
of SERS detection. 

One potential approach towards creating complex nanomorphologies in
metals is to utilize the spontaneous dewetting of thin films \cite{Herminghaus98,Khenner09,favazza06d,trice08,Rack08,Boneberg08}.
In the classical spinodal dewetting instability, an initially smooth
film is unstable to height fluctuations because attractive intermolecular
forces can exceed the stabilizing effect of interfacial tension \cite{vrij66,reiter92,Herminghaus98}.
As a result, a narrow band of wavelengths can spontaneously grow,
eventually leading to film rupture and, more importantly, to morphologies
with well-defined length scales \cite{Wyart90,sharma98-1}. One of
the important observations in polymer dewetting is the behavior of
the dewetting morphology as the film progresses from its initially
smooth state to a final stable state of particles. It has been observed
that below a transition thickness $h_{T}$, the intermediate stage
dewetting morphology consists of bicontinuous structures, while above,
it consists of regularly sized holes \cite{Xie98,Seemann01b}. This
change in morphology has been attributed to the form of the intermolecular
forces influencing dewetting \cite{sharma98-1}. Specifically, for
ultrathin films, with thickness between 1 and 20 nm, the film thickness-dependent
intermolecular forces are made up of a long range attractive component,
and a shorter-range repulsive component. The transition thickness
can be identified from the thermodynamic free energy of the system
and, as shown by Sharma and Khanna \cite{sharma98-1}, is located
at the minimum in the curvature of the free energy. Consequently,
the appearance of bicontinuous structures is correlated with films
whose initial thicknesses lie to the left of the curvature minimum,
while the formation of holes occur in films with thicknesses to the
right of the minimum. In polymers, the the magnitude of $h_{T}$ is
found to be of the order of a few nm's. More importantly, knowledge
of this transition can help guide the controlled fabrication of materials
with different morphologies and length scales.

Recently, detailed investigations by various authors have shown that
nanosecond laser irradiation of ultrathin metal films on non-wetting
substrates can initiate a similar dewetting instability, producing
robust and repeatable patterns with well-defined length scales \cite{bischof96,favazza06d,favazza06c,trice08,KrishnaPCCP09}.
Previously reported articles also showed that metal films can have
different morphologies, including bicontinuous structures or holes,
and in the case of Co on SiO$_{\text{2}}$ the transition was found
to occur between 3 to 4 nm in film thickness \cite{Favazza07a,Trice06a}.
In this work, we have explored the dewetting pathway for Ag metal
on SiO$_{\text{2}}$ substrates in the thickness range of $2\leq h\leq20$~
nm under pulsed laser melting. Besides the numerous applications associated
with Ag nanostructures, another important reason to choose Ag for
this dewetting morphological study is that nanosecond laser heating
effects do not introduce any novel dewetting effects, due primarily
to the large thermal conductivity of Ag \cite{KrishnaPCCP09}. To
capture the various morphologies between the flat film and the final
nanoparticle state, the dewetting was investigated as a function of
number of pulses. We observed that Ag shows the bicontinuous morphology
up to a thickness of \textasciitilde{}9.5~nm, while above that it
shows holes. Moreover, measurements of the characteristic length scale
showed the expected $h^{2}$ behavior associated with spinodal dewetting
over the entire thickness range investigated. Importantly, the experimentally
observed transition thickness $h_{T}$ agrees well with the value
predicted by using the sum of possible intermolecular forces for Ag
on SiO$_{\text{2}}$, and is analogous to results for polymer films.
The results from dewetting of Ag films show that various complex morphologies,
potentially useful towards plasmonic and non-linear optical properties,
can be robustly fabricated in a repeatable and controllable manner.

\section{Experimental Details}

Ag films with thickness from $\sim$2 to 20~nm were deposited in
high vacuum ($\sim1\times10^{-8}$ Torr) by pulsed laser deposition
at room temperature onto commercially obtained, optical quality, $SiO_{2}/Si$
wafers consisting of $400$ nm thick thermally grown oxide layer on
polished $Si(100)$ wafers \cite{favazza06c}. The deposition rate
was typically $\backsim0.3$ nm/min. The energy dispersive X-ray spectrometry
(EDS) was used to measure the Ag counts of the deposited films in
a scanning electron microscope (SEM). The EDS counts were converted
into an equivalent thickness value by using calibration based on step-height
measurements of the film thickness. For every film thickness we measured
the surface roughness via atomic force microscopy (AFM) and established
an upper limit of $0.5\pm0.2$ nm for the average root mean square
(RMS) roughness over the entire thickness range. Following the deposition,
the films were irradiated in vacuum by a varying number of pulses
\emph{n} from a $266$ nm ultraviolet laser having a pulse length
$\tau_{p}$ of 9 ns. Irradiation was at normal incidence by an unfocused
laser beam of area $1\times1$ mm$^{\text{2}}$ at a repetition rate
of 50 Hz. Under these vacuum deposition and irradiation conditions,
the film surface was never exposed to air and hence no role of oxygen
or an oxide layer was expected. The dewetting morphology was investigated
as a function of film thickness $h$, and the number of pulses $n$,
which typically ranged between 10 to 10,500 pulses, for irradiation
at laser energies at or just above the melt threshold of the films.
We have shown earlier that the melt threshold energy is a function
of the film thickness \cite{Trice06a}. For each thickness the melt
threshold was determined by a visible roughening of the metal film
surface, as detected under high-resolution SEM within the longest
time scale of the experiment (i.e. after $10,500$ laser pulses) \cite{Favazza06b}.
The range of $E$ used for the thickness regime investigated here
was $60\leq E\leq120$ mJ/cm$^{\text{2}}$. For this irradiation condition,
the heating and cooling rate of the metal film was of the order of
$10^{10}$ K/s with a total heating plus cooling time per pulse of
$\sim100$ ns, which was much smaller than the spacing between pulses
of $20$ ms. Consequently, as we have quantitatively shown earlier
\cite{favazza07b}, negligible contribution to the morphology evolution
was expected from processes in the solid state. Hence, any morphology
changes occurred primarily during the liquid phase following each
pulse. We also confirmed by EDS measurements that the laser irradiation
did not result in substantial evaporation of the Ag, even after the
longest irradiation experiments of 10,500 pulses.

\section{Results and discussion}

Figure \ref{fig:AFM-Roughness}(a) shows the AFM micrograph of as-deposited
2 nm Ag film on SiO$_{\text{2}}$ substrate. The average RMS roughness
of the film was measured by plotting the profile of the film along
a horizontal dashed line (see Fig. \ref{fig:AFM-Roughness}(a)), which
is shown in Fig. \ref{fig:AFM-Roughness} (b), indicating RMS roughness
to be $\sim$0.5$\pm$0.05 nm. Similar measurements were made for
all the films investigated. In Fig. \ref{fig:EarlyStageMorphology}
(a-f), a series of SEM images denoting the early stage dewetting morphology,
following irradiation by 10 laser pulses, is shown for a different
Ag film thickness. The important evidence from this series of images
is the distinct transition in the general nature of the morphology
between the 9.5 and 11.5 nm films. Up to 9.5 nm {[}Fig. \ref{fig:EarlyStageMorphology}(a-d){]},
the morphology generally consists of asymmetric undulations or a bicontinuous
type structure. On the other hand, for the 11.5 and 20 nm films {[}Fig.
\ref{fig:EarlyStageMorphology}(e-f){]}, regular holes are clearly
visible. Quantitative information about the spatial characteristics
of these morphologies was obtained by evaluating the fast Fourier
transform (FFT) of the SEM image contrast, which is related to the
dewetting film's height variations. The resulting FFT information
of the contrast correlation is shown in the inset of each figure,
and the important information here is the annular form for each of
the films. This annular FFT is indicative of a narrow band of characteristic
length scales for the height variations on the surface. This is an
important observation given that the dewetting morphology can progress
via either of the three pathways: homogeneous nucleation, heterogeneous
nucleation, or spinodal dewetting \cite{degennes03}. In the case
of homogeneous nucleation, the features are randomly distributed,
both spatially, and in time, and no characteristic length scale should
appear in this type of dewetting \cite{stange97}. Heterogeneous nucleation
can occur due to defects, impurities or other experimentally imposed
heterogeneities. In this type of dewetting, a characteristic length
scale could appear at the early stages of dewetting only in the presence
of available ordered nucleation sites. However, we have not observed
spatially ordered heterogeneities on the substrate surface, as well
as on the as-deposited films, prior to irradiation. Therefore, the
results presented here point strongly to the third option, which is
spinodal dewetting \cite{thiele98}.

The characteristic length scale associated with spinodal dewetting
is established at the very early stages of film deformation \cite{Herminghaus98}.
The early stage undulations, which occur prior to the appearance of
large height variations in the film, are extremely difficult to capture
experimentally. However, the subsequent morphology, which is a result
of ripening of the initial undulations, forms as dewetting progresses
and has length scales directly related to the initial length scale.
Therefore, as shown by many authors, the final nanoparticle length
scale, can be used as a measure of the thickness-dependent behavior
\cite{reiter92,favazza06d}. Here, the length scales were measured
at different stages of the dewetting process as a function of film
thickness. In Fig. \ref{fig:Evolution}(a-c), the progression of the
morphology is shown for the 4.5 nm film, while Fig. \ref{fig:Evolution}(e-g)
shows the same for the 11.5 nm films as a function of laser pulses
between 10 and 10,500 shots. Figures \ref{fig:Evolution}(d) and (h)
are the radial distribution functions (RDF) for each stage for the
4.5 and 11.5 nm film, respectively. From the position of the peaks
in such RDF measurements, obtained directly from the FFT's, it was
possible to generate the characteristic length scale present in the
pattern at each stage. The result of measuring length scales from
these progressions is shown in Fig. \ref{fig:PlotLengthScales}(a).
The early stage behavior is shown by the closed squares, the intermediate
stage is shown by open triangles, while the final nanoparticle state
is shown by open circles. One important observation from this measurement
is that no dramatic change in length scale is seen when the morphology
changes from the bicontinuous to the hole structures, i.e. between
9 to 11 nm. This strengthens the argument that both morphologies can
arise for spinodal dewetting of Ag. Since the characteristic length
scale for spinodal dewetting is known to vary as $\lambda_{sp}\propto h^{2}$,
we have also plotted $h^{2}$ trend lines for the early stage (solid
line) and nanoparticle stage (dotted line) data sets with an $h^{2}$
trend. The early stage dewetting agrees reasonably well with the spinodal
trend over the regime investigated, confirming a previously reported
result for Ag, based on the behavior of the nanoparticle state \cite{KrishnaPCCP09}.
An important observation can be made on the apparent deviation of
the intermediate and final stage length scales for the thickest film
investigated, i.e. the 20 nm film. The nanoparticle length appears
much smaller than the trend (dotted line), while the intermediate
state appears much larger. This can be understood as follows. The
drop in the nanoparticle length scale is due to the nearest-neighbor
interparticle spacing being dominated by a Rayleigh-like break-up
of the arms of the polygon, as shown in Fig. \ref{fig:PlotLengthScales}(b).
In contrast, particles in the 11.5 nm film form at the vertex of the
polygons formed during the dewetting stage {[}Fig. \ref{fig:Evolution}(g){]}.
A similar argument, based on a change in the feature shape being measured,
is likely to explain the intermediate state behavior. As shown in
Fig. \ref{fig:Evolution}(e-g), the progression of the hole morphology
is through merging of the holes into polygons, whose size (diameter)
will be dependent upon the number of holes it was formed from. It
is quite likely that the large increase in length scale for the intermediate
state of the 20 nm film is because the measured length scale is for
polygons formed from numerous holes.

The above results provide a strong case for Ag dewetting via the spinodal
process, with additional support coming from that fact that similar
morphological characteristics are observed when polymer films dewet
by the spinodal instability. As mentioned in the introduction, polymer
films have been observed to have a bicontinuous to hole transition
at thicknesses $h_{T}$ of a few nm and the theoretical position of
this thickness has been correlated to the nature of the free energy
\cite{sharma98-1,Seemann01b}. Specifically, the location of minima
in the free energy curvature has been correlated to this transition
thickness. To determine if a similar behavior is seen for the Ag metal,
a sum of different types of attractive and repulsive intermolecular
interactions have been used to estimate the free energy and its curvature
as a function of thickness h for Ag on SiO$_{\text{2}}$. Since
the films are very thin, we have neglected the gravitational term
in our calculations.

\subsection{Free energy analysis to determine transition thickness $h_{T}$\label{sub:Free-energy-analysis}}

The intermolecular interaction free energy of a uniform thin film
can be realized by first describing the disjoining pressure acting
on a film. Consider a liquid film of height $h$ on top of a flat
substrate. The thickness dependent disjoining pressure, $\Pi(h)$,
can be expressed by adding a long-range attractive (Van der Waals
interaction) and a short range repulsive interaction between the various
interfaces formed as a result of having the film on a substrate. The
repulsive term typically consists of two parts: (i) a Lennard-Jones
(L-J) type repulsion \cite{Mitlin94} which appears due to the electron
cloud interaction, and (ii) an electrostatic force, which is a result
of electric layers forming at the liquid-substrate interface. 

The total disjoining pressure, by considering a long-range attraction,
a short range repulsion expressed as a Lennard-Jones type form \cite{Mitlin94},
and the electrostatic force \cite{sharma98-1,Mitlin94}, is given
by:

\begin{equation}
\Pi(h)=\frac{A}{h_{c}^{3}}\left[\left(\frac{h_{c}}{h}\right)^{3}-\frac{1}{3}\left(\frac{h_{c}}{h}\right)^{9}\right]+\frac{S^{p}}{l}\, exp(-h/l)\label{eq:Potential-Mitlin}\end{equation}

\noindent Here, A is the Hamaker coefficient, which has a negative
value in units of Joule, and $h_{c}$ is the critical length at minimum
$\Pi(h)$. $S^{p}$ is the spreading coefficient which is related
to the magnitude of electrostatic part of the disjoining pressure,
and $l$ is a correlation (or Debye) length. Using Eq. \ref{eq:Potential-Mitlin},
the free energy density (energy/area) of a uniform film can be written
as:

\begin{equation}
\Delta G=\frac{A}{2h^{2}}-\frac{Ah_{c}^{6}}{24h^{8}}+S^{p}exp(-h/l)\label{eq:TotalFreeEnergy-Mitlin}\end{equation}

\noindent where, $h_{c}$ is defined as: \begin{equation}
\frac{A}{h_{c}^{2}}\frac{3^{4/3}}{8}=-2\gamma sin^{2}\left(\theta/2\right)\label{eq:Critical-length}\end{equation}

\noindent and $h_{T}$ (the transition thickness) is calculated from
the position of the minimum in the free energy curvature as:

\begin{equation}
\frac{\partial^{2}\Delta G}{\partial h^{2}}=\frac{3A}{h^{4}}-\frac{3Ah_{c}^{6}}{h^{10}}+\frac{S^{p}}{l^{2}}exp(-h/l)=0\label{eq:Cut-off-length-Mitlin}\end{equation}

The above analysis was performed for Ag on SiO$_{\text{2}}$ substrate,
and the correlation length $l$ was taken to be in the range of $0.2-1.0$
nm \cite{Sharma93}. The values of the constants used in the analysis
for Ag are given in Table \ref{tab:List-of-metal}. Table \ref{tab:List-of-metal}
also consists of material parameters for few other metals of interest,
where the value of $h_{T}$ is estimated by considering that $l$
lies between 0.2 to 1 nm. The typical transition thickness ($h_{T}$)
calculated using Eq. \ref{eq:Cut-off-length-Mitlin} for Ag appears
to be between 0.3 and 11.45 nm. The free energy and curvature plots
using Eq. \ref{eq:TotalFreeEnergy-Mitlin} \emph{and} \ref{eq:Cut-off-length-Mitlin}
and the correlation length $l=1$ nm for Ag are shown in Fig. \ref{fig:FreeEnergy}.
With this result we can make the following observations. In polymer
films, the bicontinuous structures are observed for the thinner films,
while holes are observed in the thicker ones. Similarly, for the presented
case of Ag films on SiO$_{\text{2}}$ substrate, the bicontinuous
structures are obtained for films $h_{T}\leq9.5$ nm, while above
this thickness the morphology evolution starts with the formation
of holes. This experimental result matches well with the theoretical
prediction of 11.45 nm, provided a correlation length of $l=1$ nm
is used.

\section{Conclusion}

In conclusion, we have investigated the morphology evolution in nanometer
thick Ag films $(2\leq h\leq20\, nm)$ under melting by nanosecond
laser pulses. The initial dewetting morphology is found to be a bicontinuous
morphology for films with thickness $\leq$9.5~nm, while discrete
holes appear for films with thickness > 11.5~nm. Evaluation of the
characteristic dewetting length scales indicated a good agreement
with a $h^{2}$ trend, which was evidence for the spinodal dewetting
instability. The observations of the different morphologies and a
transition thickness is consistent with the behavior observed previously
for polymer films. The experimentally observed transition value for
Ag, of between 9.5 - 11.5 nm, agrees well with predictions from intermolecular
forces. This work shows that complex and controllable morphologies
can be obtained for Ag, but more insight is needed to understand the
physical origins of dewetting pattern morphologies in thin metal films.

RK acknowledges support by the National Science Foundation through
CAREER grant DMI- 0449258, grant NSF-DMR-0856707, and grant NSF-CMMI-0855949.
M.K. acknowledges WKU awards 10-7016 and 10-7054.

\pagebreak{}

\bibliographystyle{ieeetr}

\pagebreak

\begin{center}
\begin{table}[H]
\caption{List of metal parameters required for free energy analysis\label{tab:List-of-metal}}

\centering{}\begin{tabular}{|>{\centering}m{0.5in}|>{\centering}m{0.5in}|>{\centering}m{0.5in}|>{\centering}m{0.5in}|>{\centering}m{0.75in}|>{\centering}m{0.75in}|>{\centering}m{0.75in}|}
\hline 
{\footnotesize Metal} & {\footnotesize $\gamma_{M/Vacc}$ (J-m$^{\text{-2}}$)}{\footnotesize \par}

\cite{Lu05} & {\footnotesize $\theta$ (degree)} & {\footnotesize $S^{p}$ (J-m$^{\text{-2}}$)}{\footnotesize \par}

\cite{sharma93-a} & {\footnotesize $A$ }{\footnotesize \par}

{\footnotesize (J)}{\footnotesize \par}

\cite{israelachvili92} & {\footnotesize $h_{c}$ using Eq. \ref{eq:Critical-length}}{\footnotesize \par}

{\footnotesize (nm)} & {\footnotesize $h_{T}$ using Eq. \ref{eq:Cut-off-length-Mitlin}}{\footnotesize \par}

{\footnotesize (nm)}\tabularnewline
\hline
\hline 
{\footnotesize Ag} & {\footnotesize 0.925} & {\footnotesize 82} & {\footnotesize 0.79} & {\footnotesize -1.33x10$^{\text{-19}}$} & {\footnotesize 0.300} & {\footnotesize 0.3 - 11.45}\tabularnewline
\hline 
{\footnotesize Au} & {\footnotesize 1.145} & {\footnotesize 88} & {\footnotesize 1.11} & {\footnotesize -1.76x10$^{\text{-19}}$} & {\footnotesize 0.293} & {\footnotesize 0.29 - 11.92}\tabularnewline
\hline 
{\footnotesize Co} & {\footnotesize 1.882} & {\footnotesize 101} & {\footnotesize 2.26} & {\footnotesize -3.1x10$^{\text{-19}}$} & {\footnotesize 0.273} & {\footnotesize 0.27 - 12.10}\tabularnewline
\hline 
{\footnotesize Cu} & {\footnotesize 1.304} & {\footnotesize 92} & {\footnotesize 1.35} & {\footnotesize -2.08x10$^{\text{-19}}$} & {\footnotesize 0.288} & {\footnotesize 0.29 - 11.62}\tabularnewline
\hline 
{\footnotesize Fe} & {\footnotesize 1.870} & {\footnotesize 101} & {\footnotesize 2.24} & {\footnotesize -2.9x10$^{\text{-19}}$} & {\footnotesize 0.265} & {\footnotesize 0.26 - 11.88}\tabularnewline
\hline 
{\footnotesize Mn} & {\footnotesize 1.152} & {\footnotesize 88} & {\footnotesize 1.12} & {\footnotesize -8.12x10$^{\text{-22}}$} & {\footnotesize 0.019} & {\footnotesize 2.64 - 19.23}\tabularnewline
\hline 
{\footnotesize Ni} & {\footnotesize 1.781} & {\footnotesize 100} & {\footnotesize 2.10} & {\footnotesize -9.9x10$^{\text{-20}}$} & {\footnotesize 0.160} & {\footnotesize 1.13 - 13.47}\tabularnewline
\hline 
{\footnotesize Pt} & {\footnotesize 1.746} & {\footnotesize 100} & {\footnotesize 2.04} & {\footnotesize -2.56x10$^{\text{-19}}$} & {\footnotesize 0.259} & {\footnotesize 0.26 - 11.94}\tabularnewline
\hline 
{\footnotesize Ti} & {\footnotesize 1.525} & {\footnotesize 96} & {\footnotesize 1.69} & {\footnotesize -4.32x10$^{\text{-19}}$} & {\footnotesize 0.372} & {\footnotesize 0.37 - 10.72}\tabularnewline
\hline 
{\footnotesize V} & {\footnotesize 1.770} & {\footnotesize 100} & {\footnotesize 2.08} & {\footnotesize -1.74x10$^{\text{-19}}$} & \centering{}{\footnotesize 0.212} & {\footnotesize 0.21 - 12.58}\tabularnewline
\hline
\end{tabular}
\end{table}

\par\end{center}

\pagebreak

\begin{center}
\begin{figure}[H]
\centering{}\subfloat[]{

\includegraphics[height=2in]{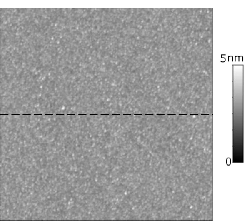}} \subfloat[]{

\includegraphics[height=2in]{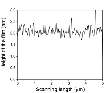}}\caption{(a) AFM image of an as deposited 2 nm Ag film on SiO$_{\text{2}}$ (image
size is 5x5 $\mu m^{2}$). (b) A line profile taken along a horizontal
dashed line of the AFM image (a) indicating the root mean square roughness
(RMS) of the Ag film is $\sim$0.5$\pm$0.05 nm. \label{fig:AFM-Roughness} }

\end{figure}

\par\end{center}

\pagebreak

\begin{center}
\begin{figure}[H]
\begin{centering}
\subfloat[]{

\includegraphics[height=2in]{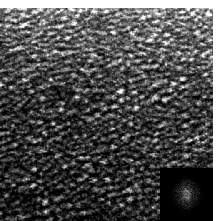}} \subfloat[]{

\includegraphics[height=2in]{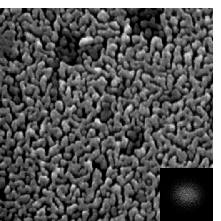}}
\par\end{centering}

\begin{centering}
\subfloat[]{

\includegraphics[height=2in]{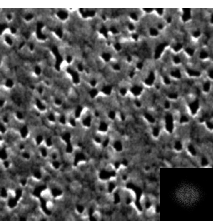}} \subfloat[]{

\includegraphics[height=2in]{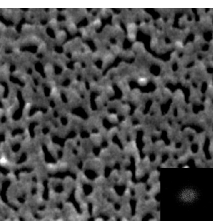}}
\par\end{centering}

\centering{}\subfloat[]{

\includegraphics[height=2in]{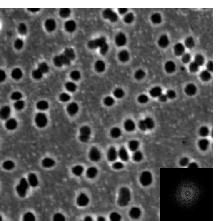}} \subfloat[]{

\includegraphics[height=2in]{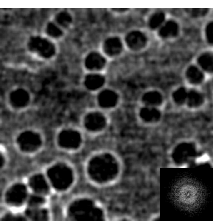}}\caption{(a-f) SEM images of the early stage dewetting morphologies following
irradiation by 10 laser pulses. The film thickness from (a) to (f),
corresponds to 2, 4.5, 7.4, 9.5, 11.5, and 20 nm, respectively. Also
shown in the inset of each figure is the FFT of the contrast correlations
of the SEM images. The annular spectrum indicates that a well-defined
length scale characterizes each pattern. The size of image (a) is
0.5x0.5 $\mu m^{2}$, images (b-d) are 1.5x1.5 $\mu m^{2}$, and images
(e) and (f) are 3x3 $\mu m^{2}$ and 10x10 $\mu m^{2}$, respectively.
\label{fig:EarlyStageMorphology} }

\end{figure}

\par\end{center}

\begin{center}
\begin{figure}[H]
\begin{centering}
\subfloat[]{

\includegraphics[width=1.5in]{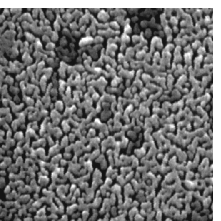}} \subfloat[]{

\includegraphics[width=1.5in]{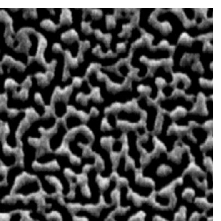}}\subfloat[]{

\includegraphics[width=1.5in]{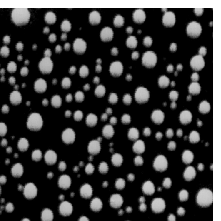}} \subfloat[]{

\includegraphics[width=1.8in]{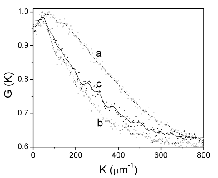}}
\par\end{centering}

\begin{centering}
\subfloat[]{

\includegraphics[width=1.5in]{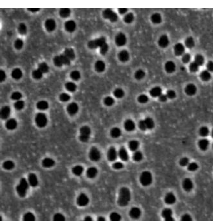}} \subfloat[]{

\includegraphics[width=1.5in]{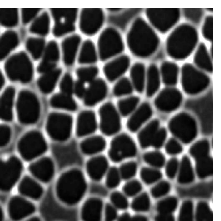}}\subfloat[]{

\includegraphics[width=1.5in]{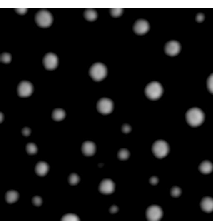}} \subfloat[]{

\includegraphics[width=1.8in]{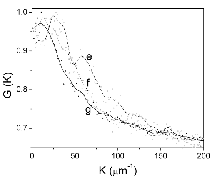}}
\par\end{centering}

\caption{(a-c) SEM images (1.5x1.5 $\mu m^{2}$) of progression of dewetting
in the 4.5 nm thick film with increasing number of laser pulses (10,
100, 10,500). (d) Plot of the radial distribution function (RDF) for
each stage of dewetting. The peak position in RDF was used to estimate
the characteristic length scales. (e-g) Progression (3x3 $\mu m^{2}$)
of dewetting in the 11.5 nm thick film with increasing number of laser
pulses (10, 100, 10,500). (h) Plot of the radial distribution function
for each stage. The letters in plots (d) and (h) indicate the RDF's
for the corresponding SEM images. \label{fig:Evolution}}

\end{figure}

\par\end{center}

\pagebreak

\begin{center}
\begin{figure}[H]
\begin{centering}
\subfloat[]{

\includegraphics[height=2.5in]{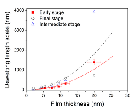}} \subfloat[]{

\includegraphics[height=2.5in]{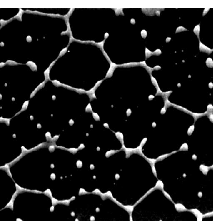}}
\par\end{centering}

\caption{(a) Plot of the characteristic length scale for various stages of
progression as a function of film thickness (\emph{h}). The early
stage behavior is shown by the closed squares, the intermediate stage
is shown by open triangles, and the final nanoparticle state is shown
by open circles. Trend lines with $h^{2}$ variation for the early
stage (solid line) and nanoparticle stage (dotted line) data are also
shown. For clarity, the error bars for the intermediate and final
state are not shown. (b) SEM image (15x15 $\mu m^{2}$) of the intermediate
stage to nanoparticle stage transition for a 20 nm film showing that
the break-up of the arms of the polygons is via a Rayleigh-type process
leading to multiple nanoparticles in each arm. \label{fig:PlotLengthScales}}

\end{figure}

\par\end{center}

\pagebreak

\begin{center}
\begin{figure}[H]
\begin{centering}
\subfloat[]{

\includegraphics[width=5in]{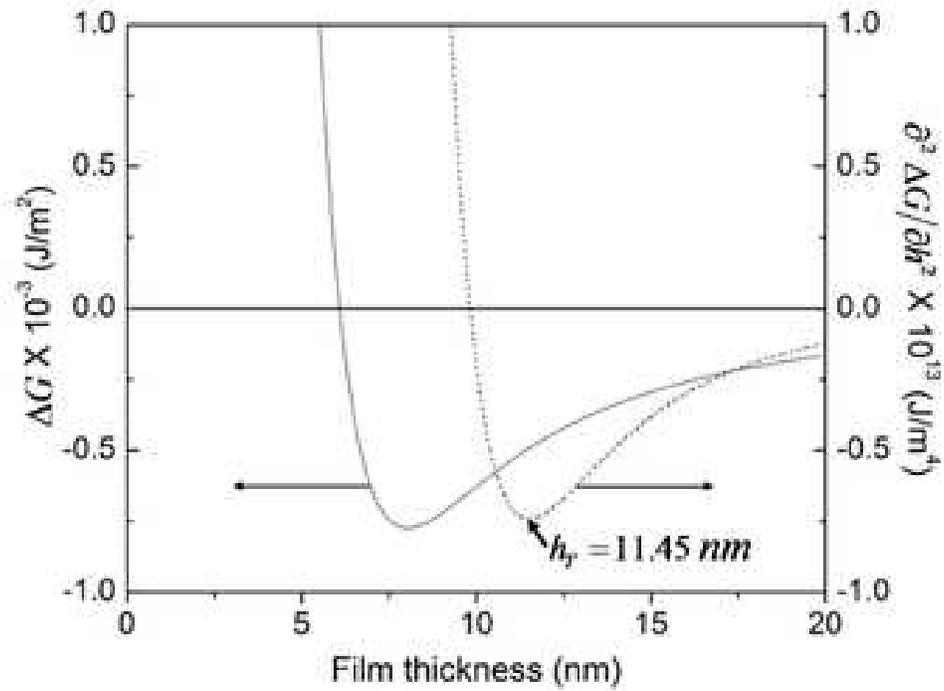}}
\par\end{centering}

\caption{Plot of the free energy, $\Delta G$, (solid line) and free energy
curvature, $\frac{\partial^{2}\Delta G}{\partial h^{2}}$, (dashed
line) vs film thickness for Ag on SiO$_{\text{2}}$ using Eq. \ref{eq:TotalFreeEnergy-Mitlin}
and Eq. \ref{eq:Cut-off-length-Mitlin}, respectively. The transition
thickness $h_{T}$ corresponds to the minimum in the curvature and
occurs at \textasciitilde{} 11.45 nm for a correlation length of 1
nm. \label{fig:FreeEnergy}}

\end{figure}

\par\end{center}
\end{document}